\documentclass{eas}
\usepackage{graphicx}
\newcommand{\hMsun}{{\ifmmode{h^{-1}M_\odot}\else{$h^{-1}M_\odot$}\fi}}
\newcommand{\hMpc}{{\ifmmode{h^{-1}{\rm Mpc}}\else{$h^{-1}$Mpc}\fi}}
\newcommand{\hkpc}{{\ifmmode{h^{-1}{\rm kpc}}\else{$h^{-1}$kpc}\fi}}

\begin{document}

\TitreGlobal{Mass Profiles and Shapes of Cosmological Structures}

\title{Halo Shapes and their Relation to Environment}
\author{Gottl\"{o}ber, S.}\address{Astrophysical Institute Potsdam, An der Sternwarte 16, 14482 Potsdam, Germany}
\author{Turchaninov, V.}\address{Keldysh Institute for Applied Mathematics,
Miusskaja Ploscad 4,  125047 Moscow, Russia}
\runningtitle{Halo Shape }
\setcounter{page}{23}
\index{Gottl\"{o}ber, S.}
\index{Turchaninov, V.}

%
\begin{abstract} 
  Using high resolution DM simulations we study the shape of dark
  matter halos. Halos become more spherical with decreasing mass. This
  trend is even more pronounced for the inner part of the halo.
  Angular momentum and shape are correlated. The angular momenta of
  neighboring halos are correlated.  
\end{abstract}

\maketitle

%
\section{Introduction}

According to the hierarchical scenario of cosmological structure
formation the backbone of galaxy formation and evolution are the halos
of cold dark matter, which emerge from a Gaussian primordial density
fluctuation field and assemble through gravitational processes. The
properties of the dark matter halos are of great interest to
understand the observed distribution of galaxies.  To study halos in
different environments we have performed a series of numerical
simulations using the new MPI version of the Adaptive Refinement Tree
code (Kravtsov et al. 1997).  In the simulations we adopt the flat
$\Lambda$CDM cosmology with ($\Omega_{\rm m} = 0.3$, $\Omega_{\Lambda}
= 0.7,$ $h = 0.7,$ and $\sigma_8 = 0.9$). We have chosen a simulation
box of $80 \hMpc$ side length. In the simulation Box80G the whole
$80\hMpc$ volume was resolved with $512^3$ equal-mass particles ($3.2
\times 10^{8}\hMsun$). The force resolution was $1.8 \hkpc$. In a
second run (Box80S) we have resimulated in the same box a spherical
volume of $10\hMpc$ radius and approximately mean cosmological density
with 150 million particles ($4.9 \times 10^{6}\hMsun$). In this case
the force resolution was $0.15 \hkpc$.

To find halos within the simulation we have used the hierarchical
friends-of-friends algorithm (Klypin et al. 1999) which bases on the
minimal spanning tree (MST) of the particle distribution.  The minimal
spanning tree of any point distribution is a unique well defined
quantity which describes the clustering properties of the point
process completely (e.g., Bhavsar \& Splinter 1996). The minimal
spanning tree of $n$ points contains $n-1$ connections. We are using a
fast MPI algorithm which calculates on 8 CPUs the MST of the $512^3$
particles within about 10 minutes. After topological ordering we cut
the MST using different linking lengths in order to extract catalogs
of friends-of-friends particle halos. Note, that cutting a given MST
is also a very fast algorithm.  We start with a linking length of 0.17
times the mean inter particle distance which corresponds roughly to
objects with the virialization overdensity $\rho/\rho_{\rm mean}\simeq
330$ at ($z=0$).  Decreasing the linking length by a factor of $2^n$
($n = $1,2,...)we get samples of objects with roughly $8^n$ times
larger overdensities which correspond to the inner part of the objects
of the first sample. With this hierarchical friends-of-friends
algorithm we can also detect substructures of halos. With our halo
finder we have detected about 90 000 halos with more than 50 particles
in Box80G and about 70 000 in the high resolution region of Box80S.
For the following analysis we have restricted ourself to halos with
more than 1000 particles.

\section{Shape and mass of halos}

\begin{figure}[h]
\centering
\includegraphics[width=6cm]{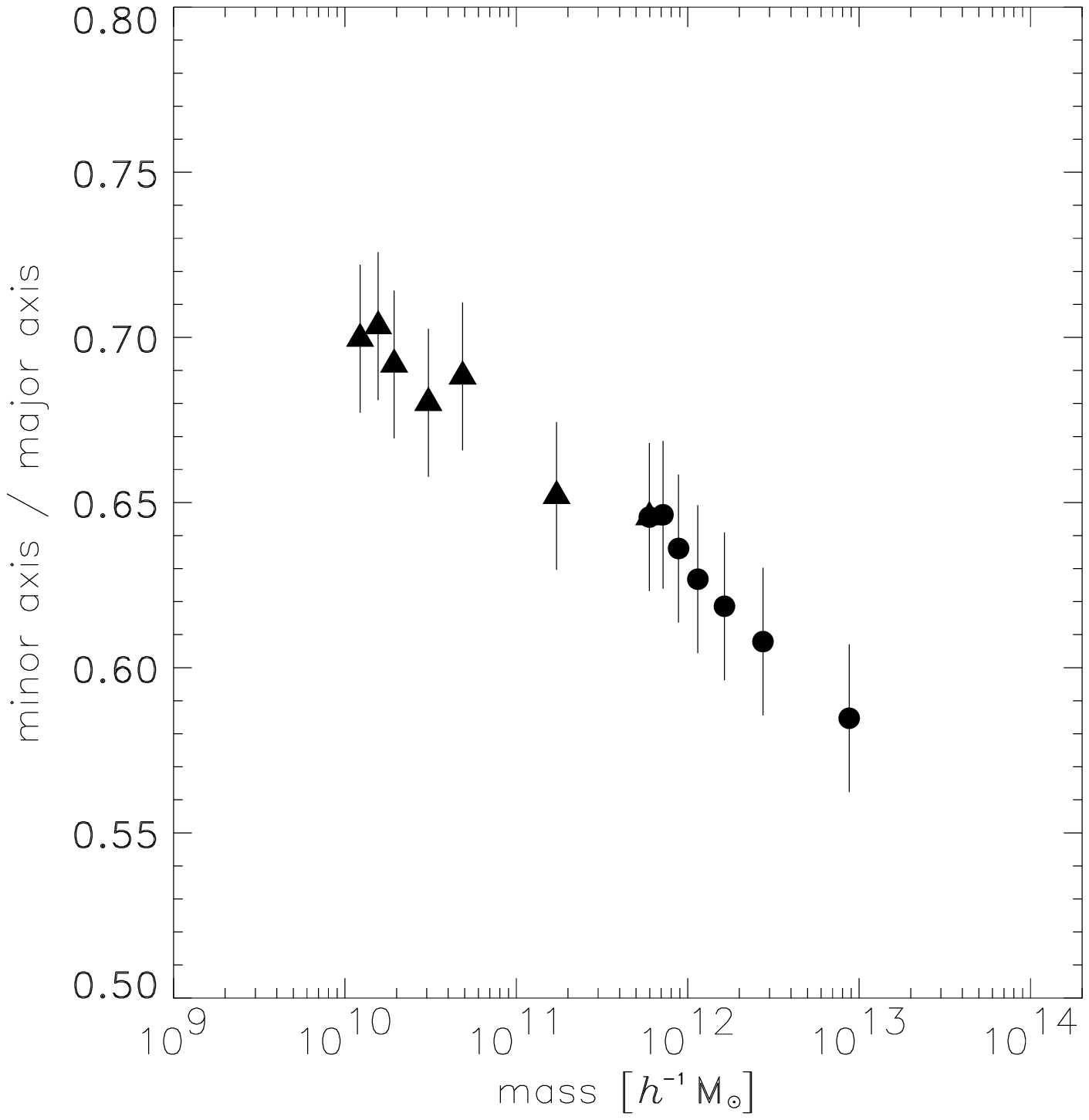}
\includegraphics[width=6cm]{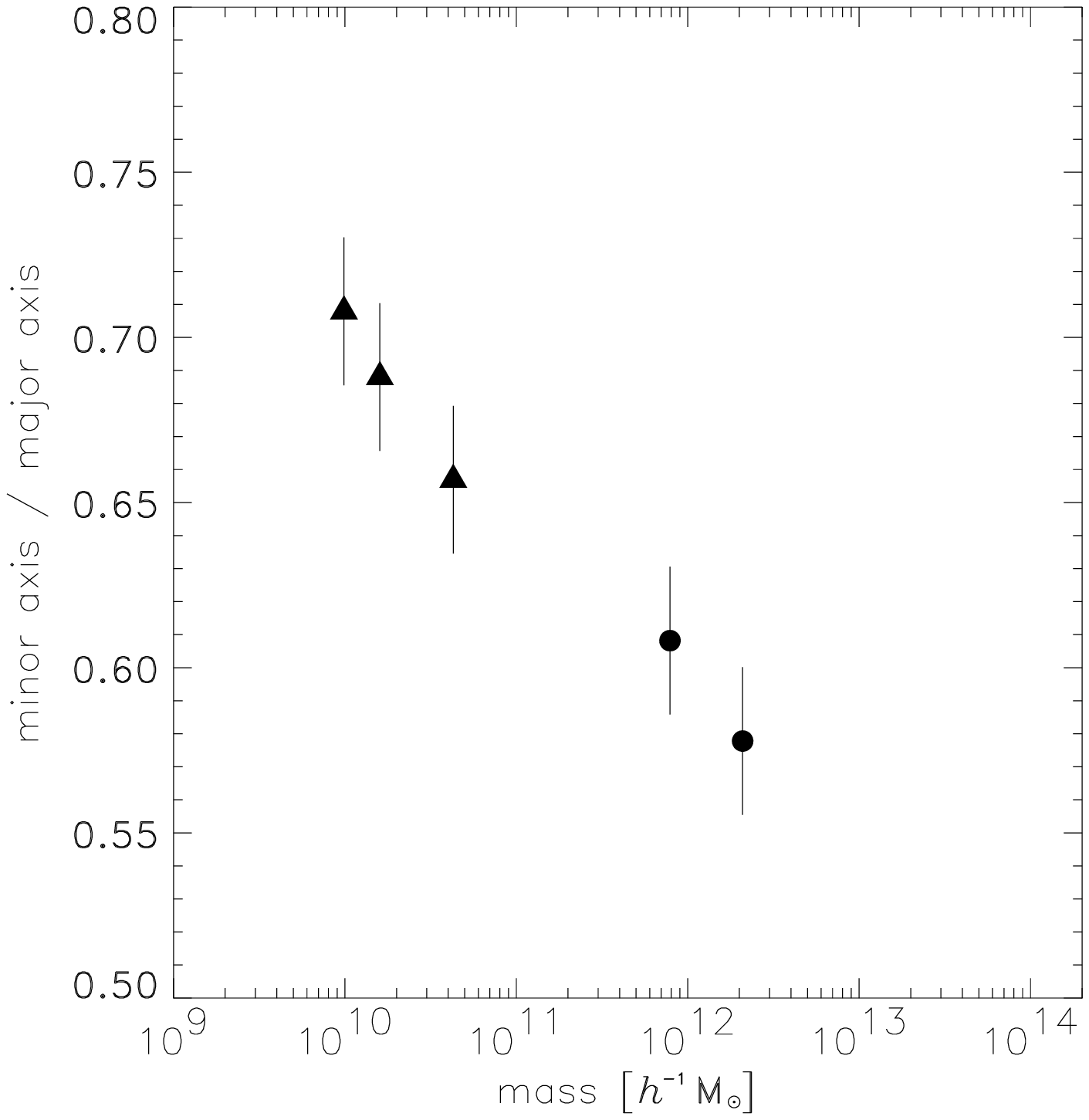}

\caption{Mean axial ratios c/a depending on mass. Left: Objects 
  with virial overdensity (relative linking length of 0.17) Right: 
  Objects with $64 \times$ virial overdensity (relative linking length
  of 0.0425) }
\label{fig_1}
\end{figure}

It is well known that dark matter halos have triaxial shapes and tend
do be prolate (e.g. Faltenbacher et al. 2002). Their shape can be
characterized by the three eigenvectors of their inertia tensor. Here
we want to study how the mean ratio of the minor axis to the major
axis depends on mass. In Fig.(\ref{fig_1}), left panel we show the
mean axial ratio $c/a$ for 500 halos per bin. Filled circles are halos
from the simulation Box80G, filled triangles are halos from Box80S.
Over three orders of magnitude we find a decreasing $c/a$ with
increasing mass of the halo. This is in accordance with Allgood et al.
(2005) who used a slightly different method for shape determination
starting from isolated spherical halos. In the right panel we show the
same for the inner part of the halo. Here we have chosen a linking
length of 0.17/4 which corresponds roughly to 64 times the virial
overdensity.  In this case the variation of shape is even steeper.

\section{Shape and angular momentum}

\begin{figure}[h]
\centering
\includegraphics[width=6cm]{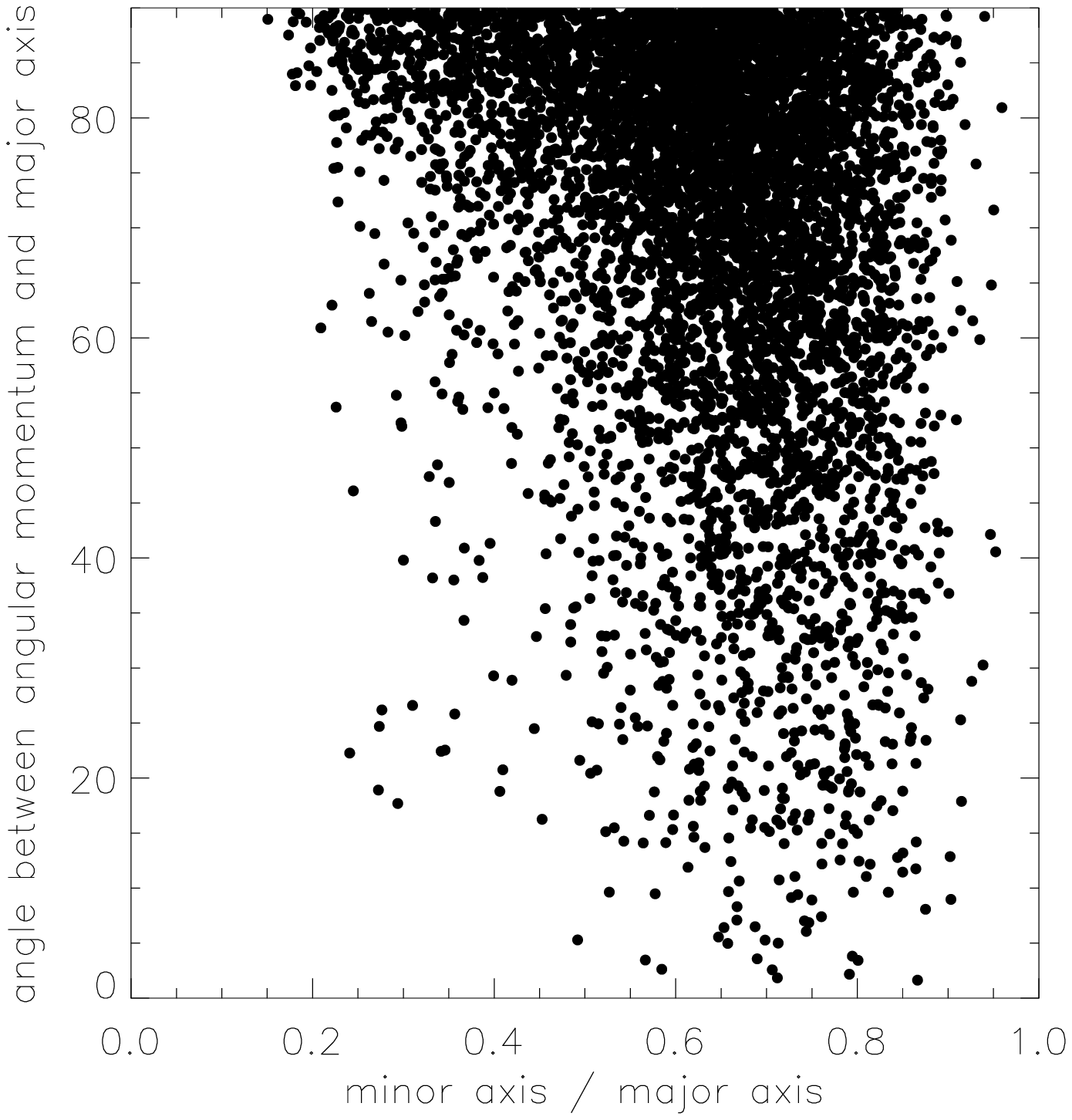}
\includegraphics[width=6cm]{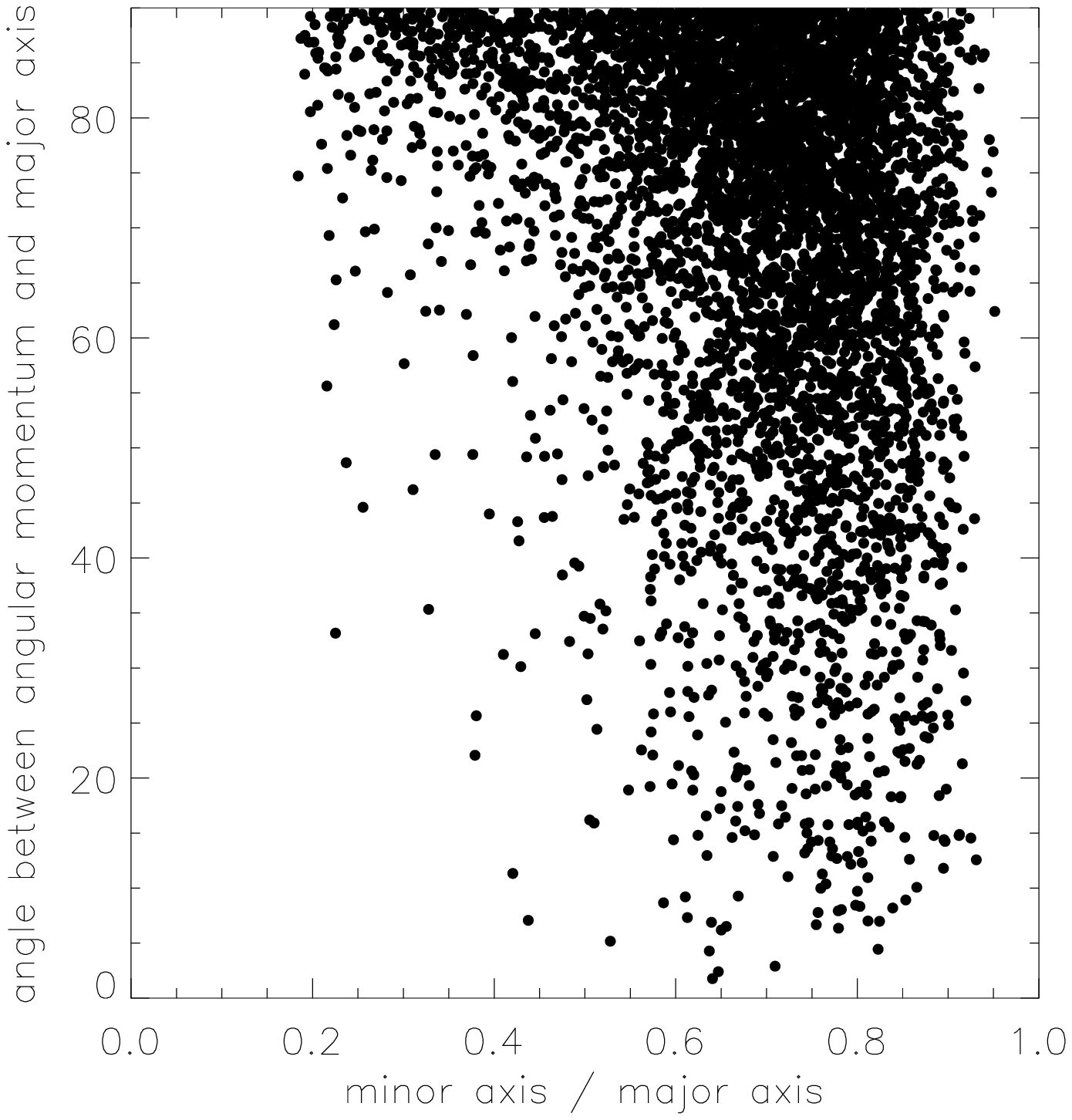}
\caption{Angle between angular momentum of dark matter halos and major 
  axis. Left: 5753 halos in the mass range $3.2 \times 10^{11} \hMsun
  \le m_{halo} \le 3.4 \times 10^{14} \hMsun $. Right: 4875 halos in
  the mass range $4.0 \times 10^{9} \hMsun \le m_{halo} \le 1.0 \times
  10^{12} \hMsun $ }
\label{fig_2}
\end{figure}

From the two simulations we have selected more than 10 000 halos with
masses of dwarf galaxies ($4.0 \times 10^{9}\hMsun $ until masses of
galaxy clusters ($3.4 \times 10^{14} \hMsun $). The left panel of
Fig.(\ref{fig_2}) shows the angle between the angular momentum and the
major axis for 5753 halos in the mass range $3.2 \times 10^{11} \hMsun
\le m_{halo} \le 3.4 \times 10^{14} \hMsun $ selected from the simulation
Box80G. For more than 76 \% of the halos the angle between the angular
momentum and the major axis is larger than $60^o$. The right panel of
Fig.(\ref{fig_2}) shows the angle between the angular momentum and the
major axis for 4875 halos in the mass range $4.9 \times 10^{9} \hMsun
\le m_{halo} \le 1.0 \times 10^{12} \hMsun $ selected from the simulation
Box80S. Again more than 75 \% of the halos show an angle larger than
$60^o$ between the angular momentum and the major axis. If the ratio
between minor axis and major axis increases the halo becomes more and
more spherical. Therefore, the major axis is not well defined and the
scatter of the angle between the angular momentum and the major axis
increases.

\section{Angular momentum of neighboring halos}

\begin{figure}[h]
\centering
\includegraphics[width=6cm]{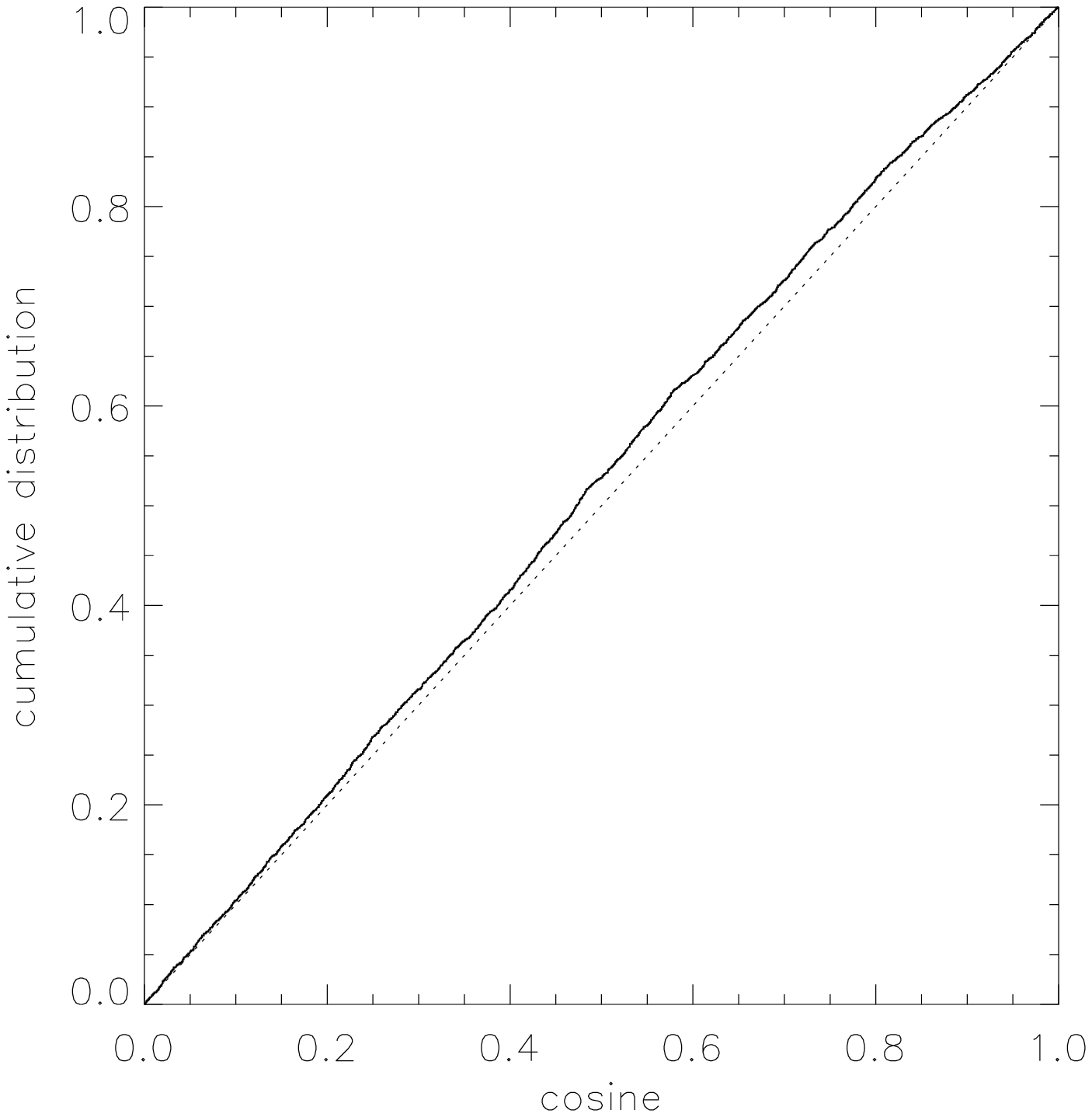}
\includegraphics[width=6cm]{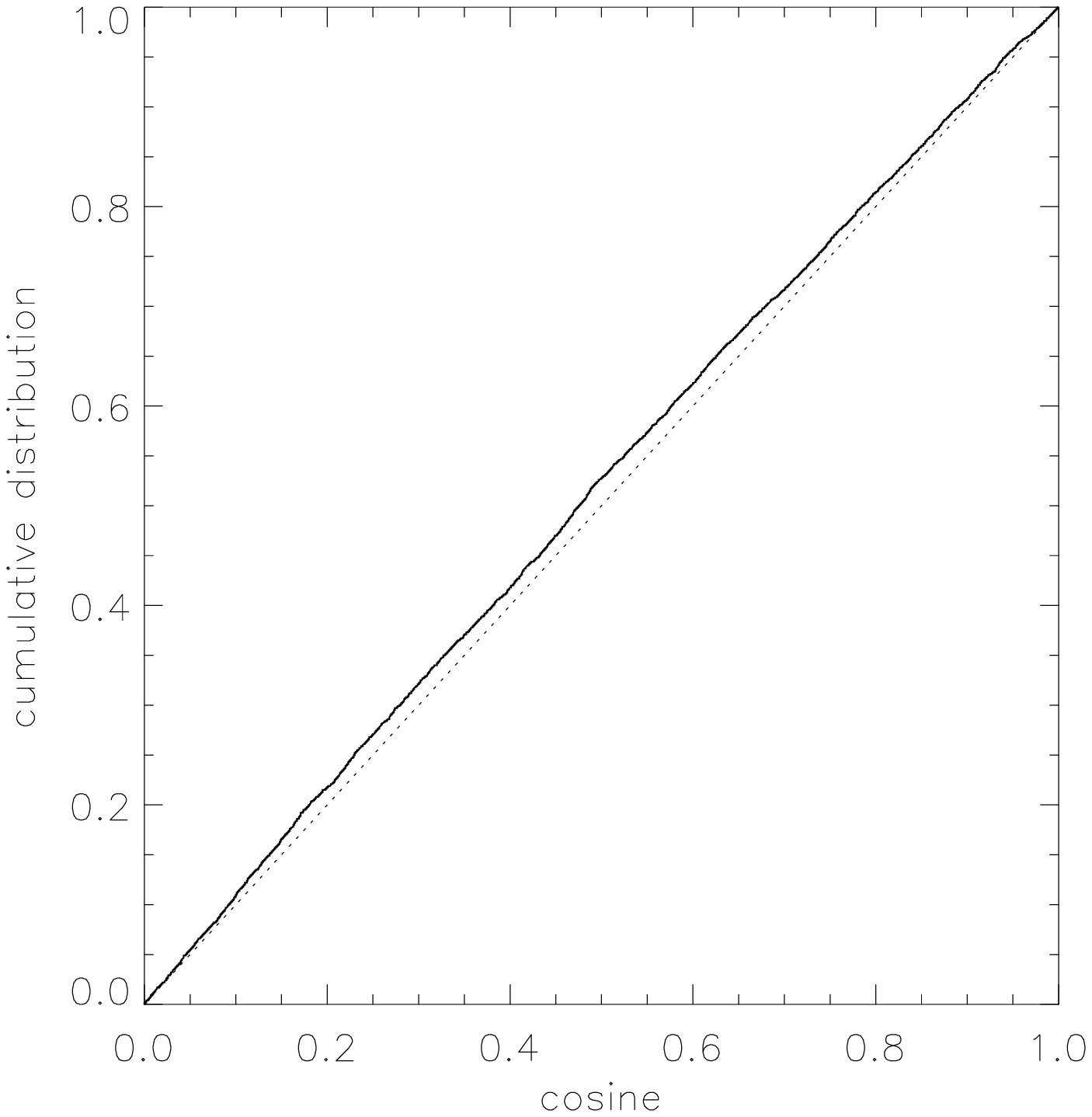}
\caption{Cumulative distribution of the absolute value of the cosine 
  between the angular momenta of two neighboring dark matter halos
  for halos in the same mass range as in Fig.(\ref{fig_2}).  }
\label{fig_3}
\end{figure}

For clusters of galaxies the major axis reflects the main infall
direction along the filaments respective the infall direction of the
last major merger (Faltenbacher et al. 2005). The shape of neighboring
galaxy clusters is correlated (Faltenbacher et al 2002). For a wide
range of masses we want to compare the angular momentum of neighboring
halos. In Fig.  (\ref{fig_3}) the cumulative distribution of the
absolute value of the cosine between the angular momenta of two
neighboring dark matter halos is shown. One can clearly see a signal
of correlation. In case of random distribution of the same number of
halos the mean value of the cosine would be 0.5 with a standard
deviation of 0.002. Here we find a mean value of 0.518. The mean to
the next 3 or 7 neighbors is 0.511 rsp. 0.509.  This is still a signal
of correlation.  The orientation of the angular momentum is correlated
with the shape of the dark matter halo (Fig. (\ref{fig_2}) as well as
with the orientation of the stellar disk (Bailin et al. 2005).
Therefore, a correlation of the angular momenta of neighboring halos
could be important for weak lensing studies where a random
distribution of galaxy shapes is assumed.

\begin{thebibliography}{}
\bibitem{} Allgood, B.F., Flores, A., Primack, J.R., Kravtsov, A.V.,
  Wechsler, R.H., Faltenbacher, A., Bullock, J.S., 2005,
  astro-ph/0508497
\bibitem{} Bailin, J., Kawata, D., Gibson, B.K., Steinmetz, M.,
  Navarro, J.F. et al., 2005, APJL 627, L17
\bibitem{} Bhavsar, S.P., \& Splinter, R.J. 1996, MNRAS 282, 1461
\bibitem{} Faltenbacher, A., Gottl\"ober, S., Kerscher, M., M\"uller,
  V., 2002, A\&A 387, 778
\bibitem{} Faltenbacher, A., Allgood, B.. Gottl\"ober, S., Yepes, G.,
  Hoffman, Y., 2005, MNRAS 362, 1099
\bibitem{} Klypin, A., Gottl\"ober, S., Kravtsov, A.~V., Khokhlov,
  A.~M. 1999, APJ 516, 530
\bibitem{} Kravtsov, A.V., Klypin, A.A., \& Khokhlov, A.M., 1997,
  ApJS, 111, 73
\end{thebibliography}
\end{document}